\documentclass[aps,prb,twocolumn,showpacs,preprintnumbers,amsmath,amssymb,superscriptaddress]{revtex4}

\usepackage{dcolumn}
\usepackage{subfigure}
\usepackage{float}
\usepackage[pdftex]{graphicx}

\newcommand{\be}{\begin{equation}}
\newcommand{\ee}{\end{equation}}
\newcommand{\bea}{\begin{eqnarray}}
\newcommand{\eea}{\end{eqnarray}}
\newcommand{\bm}[1]{\mathbf{#1}}

\newcommand{\ra}{\rangle}
\newcommand{\me}[3]{\langle #1 | #2 | #3 \rangle}

\newcommand{\lp}{\left(}
\newcommand{\rp}{\right)}
\newcommand{\ty}[1]{\mbox{\tiny #1}}

\def \cL{{\cal L}}
\def \cA{{\cal A}}
\def \cB{{\cal B}}
\def \cN{{\cal N}}
\def \cM{{\cal M}}

\def \cL{{\cal L}}

\def \xu{X_\uparrow}
\def \xd{X_\downarrow}
\def \yu{Y_\uparrow}
\def \yd{Y_\downarrow}
\def \zu{Z_\uparrow}
\def \zd{Z_\downarrow}

\begin{document}

\title{$d^0$ Perovskite-Semiconductor Electronic Structure}

\author{R. Bistritzer, G. Khalsa, and A.H. MacDonald}
\affiliation{Department of Physics, The University of Texas at Austin, Austin Texas 78712\\}

\date{\today}

\begin{abstract}

We address the low-energy effective Hamiltonian of electron doped $d^0$ perovskite semiconductors
in cubic and tetragonal phases using the $\bm{k \cdot p}$ method.
The Hamiltonian depends on the spin-orbit interaction strength, on the temperature-dependent tetragonal distortion,
and on a set of effective-mass parameters whose number
is determined by the symmetry of the crystal. We explain how these parameters can be extracted from angle resolved photo-emission,
Raman spectroscopy, and magneto-transport measurements and estimate their values in SrTiO$_3$.

\end{abstract}

\maketitle

\section{Introduction}

Transition metal oxides with perovskite structures exhibit a wide variety of interesting and
often useful effects including colossal magnetoresistance,\cite{CMR} high $T_c$ superconductivity,\cite{highTcSC}
and ferroelectricity\cite{ferroelectricity}. Correspondingly, these materials have received
intense experimental and theoretical attention for over half a century\cite{goodenough}.
Within the perovskite family, the $d^0$ materials have received particular attention,
often because of their large band gaps.
SrTiO$_3$, for example, is perhaps the most common substrate for the epitaxial growth of
oxide materials. Recently there has been growing interest in the transport properties of lightly electron doped $d^0$ perovskites.\cite{HighMobilitySTO}
In KTaO$_3$, for example, strong spin-orbit (SO) coupling facilitates
electrical manipulation of spin in a field effect transistor geometry.\cite{KTO}
The two-dimensional electron systems which form at interfaces between $d^{0}$ materials\cite{LAO_STO_Hwang} show
intriguing magnetic phases\cite{LAO_STO_magnetic} and peculiar magneto-transport features.\cite{Dagan,LiorKlein} Advanced
epitaxial growth techniques
enable $\delta$-doping of oxides\cite{deltaDopedSTO} and the fabrication of oxide hetrostructures.\cite{hetrostructures}
These relatively recent rapid advances could, it is hoped, eventually lead to useful oxide based nano-electronic devices.\cite{enterOxides}

The low-energy band structure of an oxide provides a starting point for understanding not only its bulk transport
characteristics but also
its electronic properties near $\delta$-doped layers and near
interfaces.  First principles electronic structure theory methods\cite{Mattheiss1,Mattheiss2,KTOnumerics}
are usually efficient for determining the gross structure of a band, but are not sufficiently accurate
to nail down the fine features that determine the electronic properties of the states at the bottom of the
conduction band that are important in weakly doped bulk materials, and in low-carrier-density
two-dimensional electron systems.  In particular, it appears that at present
bulk band structures in $d^{0}$ perovskites are not known accurately enough to predict the two-dimensional
bands of $\delta$-doped oxides or of interface-localized bands in oxide based hetrostructures.
This paper is primarily motivated by the goal of assisting progress in this direction.

The $\bm{k \cdot p}$ method\cite{Dresselhaus,cardona} offers an alternative and a potentially more accurate route for
characterizing band structure
near the conduction band minimum.
The method provides an effective Hamiltonian that depends on
a set of phenomenological parameters which can be small in number
when band extrema occur at high-symmetry points in momentum space.
The utility of this method hinges on the ability to extract accurate
parameter values  from experiments.  In the case of perovskites the most valuable experimental probes are
angle resolved photo-emission (ARPES), Raman spectroscopy, and magneto-transport measurements.

Many of the most studied oxides have conduction-band minima located at the center of the Brillioun zone.
We therefore apply the $\bm{k \cdot p}$ method to obtain an effective low energy Hamiltonian near the $\Gamma$ point.
At high temperatures, perovskites typically have cubic symmetry.
As the temperature is decreased the symmetry is usually lowered, most commonly to either orthorhombic or
tetragonal. The distortion can be driven by the motion of atoms along one of the cubic axes ({\em e.g.} in BaTiO$_3$)
or by a rotation of the oxygen octahedras ({\em e.g.} in SrTiO$_3$).
Structural phase transitions can also be induced by applied stress.\cite{expitaxialStressSTO}

In this work we focus on the cubic and tetragonal phases.
In section \ref{sec_low_energy_theory} we briefly describe the $\bm{k \cdot p}$ method and then use it to derive the low energy effective
theory of a $d^0$ perovskite in the vicinity of the $\Gamma$ point. In section \ref{sec_experimental_methods} we elaborate on experimental methods
for obtaining the parameters of the $\bm{k \cdot p}$ Hamiltonian. Using the experimental data accumulated over the past few decades we then study
the effective Hamiltonian of the conduction bands of SrTiO$_3$ in Section \ref{sec_STO}.
We summarize in section \ref{sec_summary}.

\section{Low energy theory  \label{sec_low_energy_theory}}

For many perovskites of current interest such as SrTiO$_3$ the conduction band minima is at the Brillouin-zone center $\Gamma$-point.
For momenta near the $\Gamma$-point the crystal field splits the ten $d$-bands into four high energy $e_g$ bands, and six lower energy $t_{2g}$ bands.
Because the crystal field induced gap is typically a few eV's, it is sufficient to
consider the $t_{2g}$ bands when constructing a low energy theory of weakly-doped $d^{0}$ materials.
In the cubic phase the $t_{2g}$ bands are degenerate at the $\Gamma$-point if
spin-orbit interactions are neglected, but are
weakly-split by typical tetragonal or orthorhombic distortions and by weak spin-orbit interactions.
Unless the Fermi energy is large compared to these splittings, spin-orbit and distortion related band parameters
must be accurately known in order to achieve a reasonable description of electronic properties.

\subsection{Effective Hamiltonian}

The unperturbed Hamiltonian in the $\bm{k \cdot p}$ perturbation theory\cite{Dresselhaus,cardona} is
\be
H_0 = \frac{p^2}{2m} + V(\bm{r}) + \frac{\hbar}{4m^2 c^2} \lp \nabla V \times \bm{p} \rp \cdot \sigma.      \label{H0}
\ee
$H_0$ consists of three terms: the kinetic energy term, the lattice potential term $V(\bm{r})$, and the spin-orbit term
($\sigma$ is the Pauli matrix vector).  The $\bm{k \cdot p}$ Hamiltonian, which acts on the periodic part of the
Bloch state, includes a second term which accounts for the
dependence of band wavefunctions on Bloch wavevector $\bm{k}$:
\be
H_{\bm{k \cdot p}} = \frac{\bm{k}}{m} \cdot \lp \bm{p} +  \frac{1}{4m^2 c^2} \bm{\sigma} \times \nabla V \rp \equiv \frac{\bm{k}}{m} \cdot \bm{P}.       \label{H_kp}
\ee
The $\bm{k \cdot p}$ method exploits the high symmetry at the $\Gamma$ point to classify the $\bm{k}=0$ wave functions
by irreducible representations (irreps) of the appropriate point group symmetry. It then uses
perturbation theory
\be
h_{ij} = \delta_{ij}k^2 + \sum_\alpha \frac{\me{\psi_i}{H_{\bm{k \cdot p}}}{\phi_\alpha} \me{\phi_\alpha}{H_{\bm{k \cdot p}}}{\psi_j}}{E_i(0)-E_\alpha(0)}      \label{h_ij}
\ee
to evaluate $t_{2g}$ projected Hamiltonian corrections to second order in the Bloch wavevector $\bm{k}$.
Hereafter we use units in which $\hbar=2m=1$ where $m$ is the bare mass of the electron.
The six $t_{2g}$ band energies $\epsilon(\bm{k})$ then follow from the secular equation
\be
\det[h_{\ty{SO}}+h_{\ty{L}}+h(\bm{k})-\epsilon(\bm{k}) I] = 0       \label{secular_eq}.
\ee
In Eq.(\ref{h_ij}) $\{ | \psi_j \ra \}$  label a basis set for the $t_{2g}$ bands and $\phi_\alpha$ is summed over bands outside the
$t_{2g}$ manifold.
The first order term was omitted in Eq.(\ref{h_ij}) since it vanishes for the perovskite structure by inversion symmetry.
The matrices $h_{\ty{L}}$ and $h_{\ty{SO}}$ account phenomenologically for tetragonal distortion and SO interactions at the $\Gamma$ point
and are discussed more explicitly below.

The wave functions at the zone center
have no covalent character
and can be spanned by the $t_{2g}$ basis
\be
\{ \xu , \yu, \zu , \xd, \yd , \zd \}.      \label{basis_functions}
\ee
Here $X,Y$ and $Z$ correspond respectively to the $|yz\ra,|xz\ra$ and $|xy\ra$ $t_{2g}$ orbitals.
Below we obtain the Hamiltonian matrix in this basis.

The lattice term $h_{\ty{L}}$ is non zero in the tetragonal phase.
If we choose a convenient zero of energy and set the $\bm{\hat{z}}$ axis along the tetragonal axis
then $h_{\ty{L}}$ has a single non-zero
matrix element:
\be
\me{Z\alpha}{V}{Z \alpha} = \Delta_{\ty T},     \label{DeltaT}
\ee
where $\alpha$ accounts for the spin. The SO term in the Hamiltonian is
\be
\lp h_{\ty{SO}} \rp_{i\alpha,k\beta} = \me{\xi_i \alpha}{\Lambda \cdot \sigma}{\xi_k \beta} = \me{\xi_i}{\Lambda_j}{\xi_k} \cdot \me{\alpha}{\sigma_j}{\beta},  \label{hSO_me}
\ee
where $\Lambda \propto \nabla V \times \bm{p}$ and $\xi_i$ is one of the  orbital basis functions.
Because $\Lambda$ transforms as a pseudovector, $\me{\xi_i}{\Lambda_j}{\xi_k} \propto \epsilon_{ijk}$ where $\epsilon_{ijk}$
is the third rank antisymmetric tensor. For example, $\me{X}{\Lambda_z}{X}$ and $\me{X}{\Lambda_z}{Z}$ vanish under reflection off the x-z plane.
Furthermore, since the matrix elements (\ref{hSO_me}) must be imaginary
\be
\me{\xi_i}{\Lambda_j}{\xi_k} = -i \frac{\Delta_{\ty{SO}}}{3} \epsilon_{ijk}.       \label{L_me}
\ee
Strictly speaking, SO coupling is described by two parameters in the tetragonal phase.
However we neglect this small correction since it is of order of $\Delta_{\ty T}$ over the band gap compared to the
spin-orbit coupling term we retain.

The $\bm{k}$-dependent part of the Hamiltonian $h$ is obtained using Eq.(\ref{h_ij}).
We show in the appendix \ref{app_H} that
\be
h =
\left(
  \begin{array}{cc}
    h_{\uparrow} & 0 \\
    0 & h_{\downarrow}     \\
  \end{array}
\right)
\ee
with
\begin{widetext}
\be
h_{\alpha} = \lp
\begin{array}{ccc}
  \cL_5 k_x^2 + \cM_5^\parallel k_y^2 + \cM_5^\perp k_z^2 &  \cN_5 k_x k_y & \cN_{45}^\star k_x k_z \\
  \cN_{5}k_x k_y & \cL_5 k_y^2 + \cM_5^\parallel k_x^2 + \cM_5^\perp k_z^2 & \cN_{45}^\star k_y k_z \\
  \cN_{45} k_x k_z & \cN_{45} k_y k_z  &  \cM_4(k_x^2+k_y^2) + \cL_4 k_z^2
\end{array}
\rp
|\alpha\ra.                      \label{h}
\ee
\end{widetext}
In the tetragonal phase the $h$ matrix depends on eight real parameters (only $\cN_{45}$ may be complex).
In the cubic phase parameter values become
independent of their subscript labels ({\em e.g.} $\cL_4=\cL_5 \to L$) and $h$ then depends on only three parameters.
The energy dispersion relations follow from Eqs.(\ref{secular_eq},\ref{DeltaT}---\ref{h}).
Because the Hamiltonian is time-reversal invariant and has inversion symmetry it gives rise to three doubly-degenerate
bands.

In the next section we discuss zone-center wave functions and energies.
The wavefunctions play a crucial role in matrix-element considerations
which powerfully expand the ability of ARPES experiments to determine the parameters of the
$\bm{k \cdot p}$-Hamiltonian.
The zone-center energies can be compared with  $t_{2g}$
band-splitting values obtained by Raman spectroscopy.

\subsection{Zone center energies and wave-functions     \label{sec_zone_center}}

The Hamiltonian at the zone center is $h_{\ty L}+h_{\ty{SO}}$. The energies are therefore
\bea
\epsilon_6 &=&  0          \nonumber \\
\epsilon^{(a)}_7 &=&  \frac{\Delta_{\ty{SO}}}{2} + \frac{\Delta_{\ty{T}}}{2} - \frac{Q}{3}      \nonumber \\
\epsilon^{(b)}_7 &=&  \frac{\Delta_{\ty{SO}}}{2} + \frac{\Delta_{\ty{T}}}{2} + \frac{Q}{3},  \nonumber \\
\label{E_tetragonal}
\eea
where
\be
Q = \frac{3}{2}\sqrt{\Delta_{\ty{SO}}^2 - \frac{2}{3} \Delta_{\ty{SO}}\Delta_{\ty{T}} + \Delta_{\ty{T}}^2}.  \label{Q}
\ee
(Energy has been shifted so that $\epsilon_6$ will vanish.)
In the cubic phase the $t_{2g}$ bands transform as $\Gamma^+_{25}$ in the absence of spin-orbit coupling.
SO interactions split the bands to $\Gamma_7^+ + \Gamma_8^+$.
When there is a tetragonal transition, the four-fold degenerate $\Gamma_8$ states further
split to $\Gamma_7+\Gamma_6$. The notation in Eqs.(\ref{E_tetragonal}) correspond to these latter irreps.

The (unnormalized) wave functions corresponding to the energies (\ref{E_tetragonal}) are
\bea
\psi^{6}_1 &=&   \xd - i\yd        \nonumber \\
\psi^{6}_2 &=&  \xu + i\yu         \nonumber \\
\psi^{7a}_1 &=&   (Q+D)\xu - i(Q+D)\yu +2\Delta_{\ty{SO}}\zd         \nonumber \\
\psi^{7a}_2 &=&  \Delta_{\ty{SO}}\xd + i\Delta_{\ty{SO}}\yd - (Q-D)\zu          \nonumber \\
\psi^{7b}_1 &=&  (Q-D)\xu - i(Q-D)\yu - 2\Delta_{\ty{SO}}\zd         \nonumber \\
\psi^{7b}_2 &=&   \Delta_{\ty{SO}}\xd + i\Delta_{\ty{SO}} \yd + (Q+D)\zu             \label{zone_center_wf}
\eea
where
\be
D = 3\Delta_{\ty T}/2-\Delta_{\ty{SO}}/2.        \label{D}
\ee
It is interesting to follow the evolution of the bands as the ratio between $\Delta_{\ty{T}}$ and $\Delta_{\ty{SO}}$ is varied from zero to infinity.
The two limits are given in table \ref{tab:wfs}.
In the cubic phase the states $\{ \psi^6 , \psi^{7a} \}$ are degenerate and are spilt off from the $\{ \psi^{7b} \}$ states by  an energy of $\Delta_{\ty{SO}}$. In the
tetragonal phase when $|\Delta_{\ty{SO}}|>|\Delta_{\ty{T}}|$ the states group to the three doubly degenerate pairs $\psi^6,\psi^{7a}$ and $\psi^{7b}$.
As the temperature is lowered the four $\psi^7$ states mix.
If eventually $|\Delta_{\ty{SO}}| \ll |\Delta_{\ty{T}}|$ then the $\psi^{7a}_1$ and $\psi^{7b}_1$ states combine to give the $\zd$ state which is purely tetragonal in character.
\begin{table}
\begin{tabular}{|c|c|}
  \hline
  &  \\
  $\Delta_{\ty T}=0$ &  $\Delta_{\ty{SO}}=0$ \\
  \hline
  &  \\
  $\lp \xd - i\yd \rp , \Gamma_6(\Gamma_8)$ &  $\lp \xd - i\yd \rp , \Gamma_6(\Gamma_5)$ \\
   &  \\
  $\lp \xu + i\yu \rp , \Gamma_6(\Gamma_8)$ &  $\lp \xu + i\yu \rp , \Gamma_6(\Gamma_5)$ \\
  &  \\
  \hline
   &   \\
  $\left[ \xu - i\yu + 2\zd \right] , \Gamma_7(\Gamma_8)$ &  $  \zd , \Gamma_7(\Gamma_4)$ \\
  &  \\
  $\left[ \xd + i\yd - 2\zu \right] , \Gamma_7(\Gamma_8)$ &  $ -\zu , \Gamma_7(\Gamma_4)$  \\
  &   \\
  $\left[ -\xu + i\yu + \zd  \right] , \Gamma_7 $   & $-\lp \xu - i \yu \rp , \Gamma_7(\Gamma_5)$ \\
   &  \\
  $\left[ \xd + i\yd  + \zu \right] , \Gamma_7$ &  $\lp \xd + i \yd \rp ,  \Gamma_7(\Gamma_5)$  \\
  &  \\
  \hline
\end{tabular}
\caption{Zone center wave functions in the cubic phase with SO interactions (left column) and in the tetragonal phase in the absence of SO interactions (right column).}
\label{tab:wfs}
\end{table}

In the following
section we discuss energy dispersion relations along symmetry lines and planes,
which can be directly related to ARPES measurements and
enable some qualitative insights into the relationships between
Hamiltonian parameters and the field-orientation dependence of
magnetoresistance-oscillation frequencies.

\subsection{Energy dispersion relations for high-symmetry lines and planes  \label{sec_dispersion_relations}}

In general Eq.(\ref{secular_eq}) must be diagonalized numerically. However, simple energy dispersion relations exist
along high symmetry directions and in high-symmetry planes.

When the tetragonal distortion is large and SO interactions can be neglected,
the $t_{2g}$ bands split into $\Gamma_4+\Gamma_5$ bands.  In this limit (to order $k^4/\Delta_{\ty{T}}$)
\bea
\epsilon_4(\bm{k}) &=& \Delta_{\ty T} + \cM_4 k_\parallel^2 + \cL_4 k_z^2     \nonumber \\
\epsilon_{5\pm}(\bm{k}) &=& \cB_+ k_\parallel^2 + \cM_5^\perp k_z^2           \label{Ek_tetragonal}  \\
&\pm& \sqrt{\cB_- k_\parallel^4 - 4\left[ \cB_-^2-\cN_5^2 \right] k_x^2 k_y^2}  \nonumber
\eea
where $k_\parallel^2=k_x^2+k_y^2$ and $\cB_\pm = (\cL_5\pm\cM_5^\parallel)/2$.  To leading order in $\Delta_{\ty{SO}}$,
the $\Gamma_4$ energies remain unchanged
whereas the $\Gamma_{5\pm}$ energies vary linearly in opposite directions.
The energies (\ref{Ek_tetragonal}) are valid for any value of $\Delta_{\ty{T}}$ (but still neglecting $\Delta_{\ty{SO}}$) in the $k_z=0$ plane.
Similarly for the $k_y=0$ plane
\bea
\epsilon_{5-}(\bm{k}) &=& \cM_5^\parallel k_x^2 + \cM_5^\perp k_z^2     \nonumber \\
\epsilon_{4,5+}(\bm{k}) &=& \frac{\Delta_{\ty T}}{2} + \frac{\cM_4+\cL_5}{2}k_x^2 + \frac{\cL_4+\cM_5^\perp}{2} k_z^2           \nonumber   \\
&\pm& \frac{1}{2} \bigg( \left[ \Delta_{\ty T} + \lp \cM_4-\cL_5 \rp k_x^2 + \lp \cL_4-\cM_5^\perp \rp k_z^2 \right]^2   \nonumber  \\
&&  + 4|\cN_{45}|^2 k_x^2 k_z^2  \bigg)^{1/2}.               \label{Ek_ky0}
\eea

The $\bm{k \cdot p}$ Hamiltonian for the $t_{2g}$ bands in the cubic phase is identical to that of the valence
band p-states of zinc-blende type semiconductors.\cite{Dresselhaus,cardona}
In the presence of moderate SO interactions the dispersion relations along the three equivalent principle axes are
\bea
\epsilon_7(k) &=&   M k^2   \nonumber  \\
\epsilon_{8\pm}(k) &=&    \cB_+ k^2 + \frac{\Delta_{\ty{SO}}}{2}        \label{Ecubic_100} \\
&\pm& \sqrt{\cB_-^2 k^4 + \lp \frac{\Delta_{\ty{SO}}}{2} \rp^2 - \frac{\Delta_{\ty{SO}}}{3}\cB_- k^2}.  \nonumber
\eea
For strong SO interactions the $\psi^7$ and $\psi^8$ states can be approximately decoupled to order $k^4/\Delta_{\ty{SO}}$.
The energy dispersions are then
\bea
\epsilon_7(k) &=& \Delta_{\ty{SO}} + A k^2      \label{E_SO}  \\
\epsilon_8(\bm{k}) &=& Ak^2 \pm \sqrt{B^2k^4 + C^2(k_x^2k_y^2+k_x^2k_z^2+k_y^2k_z^2)}  \nonumber
\eea
where $ A= 1 + (L+2M)/3$, $B = (L-M)/3$, and $C^2 = \left[ N^2 - (L-M)^2 \right]/3$.
Expressions (\ref{E_SO}) were obtained by Dresselhaus et. al.[\onlinecite{Dresselhaus}].

ARPES measurements are frequently set to measure the energy dispersion in the $k_x-k_y$ plane.
For $k_z=0$ the dependence of band energies on
momenta is similar in the dominant tetragonal-splitting and
dominant spin-orbit coupling limits(compare Eqs.(\ref{Ek_tetragonal}) and (\ref{E_SO})).
One way to determine which of the two interactions is dominant
is to probe the dispersion relation along $\bm{\hat{z}}$.
A second way is to monitor the evolution of the bands as a function of temperature.
Additional methods are explained in section \ref{sec_experimental_methods} below.

The parameters of the effective Hamiltonian in the tetragonal phase are temperature dependent.
As $T$ is lowered the tetragonal distortion increases and the energy bands change accordingly.
For some crystals, such as SrTiO$_3$, the deformation is well described by a simple order parameter\cite{expitaxialStressSTO}.  It is then possible
to express the temperature dependence of the different Hamiltonian parameters via a single temperature dependent order parameter.

\section{Experimental methods for determining Hamiltonian parameters  \label{sec_experimental_methods}}

The utility of the $\bm{k \cdot p}$ method depends on the ability to extract accurate values for the Hamiltonian parameters from experiments.
ARPES, magneto-transport, and Raman spectroscopy measurements are three of the most useful experimental probes for band parameters.
In this section we focus on the ways in which these techniques can be exploited for $d^0$ perovskites with an emphasis on experimental
signatures of the tetragonal distortion.

\subsection{Raman spectroscopy      \label{sec_Raman}}

Raman spectroscopy is routinely used to measure the spectra of solids\cite{cardona}. For a low doped $d^0$ perovskite Raman spectra
can determine the band gaps at the zone center.
As explained in section \ref{sec_dispersion_relations} distinguishing between $\Delta_{\ty T}$ and $\Delta_{\ty{SO}}$ using ARPES measurements
may prove difficult. The band gaps depend both on SO interactions
and on the tetragonal distortion. Spectroscopically monitoring the energy gaps as a function of temperature
and comparing with Eqs.(\ref{E_tetragonal}) provides in principle sufficient information
to determine $\Delta_{\ty{SO}}$ and $\Delta_{\ty T}$.

\subsection{ARPES       \label{sec_ARPES}}
Angle Resolved Photoemission Spectroscopy (ARPES) has now been developed into a widely applicable
experimental tool for the measurement of bulk and surface electronic states.\cite{ARPES_review}
In a typical measurement incident monochromatic radiation excites electrons in occupied crystal states
and unbinds them from the crystal.
In the {\em sudden approximation} electrons are promoted directly from a crystal state to a vacuum plane wave state.
In this approximation the intensity of the ARPES signal associated with in-plane electron momentum $\bm{k_\parallel}$ and energy $\omega$ is
\begin{equation}
I\left(\bm{k_\parallel},\omega \right) \propto \sum_n \vert M_{\bm{p},n\bm{k_\parallel}}\vert^2 \cA_n \left( \bm{k_\parallel},\omega \right) f(\omega).       \label{I}
\end{equation}
Here the z-axis is set perpendicular to the sample's surface and we assume that the photon energy is calibrated to probe the $k_z=0$ plane.
$\cA_n$ is the electron spectral function of band $n$, $f$ is the
Fermi distribution function, and
\begin{equation}
M_{\bm{p},n\bm{k_\parallel}} \propto \langle \bm{p}  \vert \bm{A} \cdot \bm{p}  \vert \Psi_{n\bm{k_\parallel}} \rangle =    \bm{A \cdot p}  \langle \bm{p}  \vert \Psi_{n\bm{k_\parallel}} \rangle
\label{mfi}
\end{equation}
gives the probability amplitude for an electron in an initial state $\Psi_{n\bm{k_\parallel}}$ to transition to a plane wave state $\bm{p}$ via
a photon field $\bm{A}$.
The photo-emitted electrons are selectively collected according to their
emission angle and energy. Therefore in a given measurement the outgoing momentum $\bm{p}$ in Eq.(\ref{I}) is fixed
by the position of the detector and by the energy of the incoming photon.  The component of the momentum parallel to the
surface must equal the momentum of the initial state to within a surface reciprocal lattice vector.

In principle with sufficient ARPES data the occupied energy bands can be accurately mapped. The $\bm{k \cdot p}$ Hamiltonian parameters can then be determined using
the dispersion relations in section \ref{sec_dispersion_relations}.
In practice, however, experimental limits on energy and momentum resolution combined with the relatively
large number of Hamiltonian parameters and the possibility of surface states that obscure bulk bands, often
complicate comparisons between theory and experiment.

As we now explain, additional band structure information can sometimes be drawn from systematics in the
dependence of the ARPES matrix elements on the surface reciprocal lattice vector added to the transverse momentum.
Matrix elements contributions from particular $t_{2g}$ orbitals frequently vanish at particular reciprocal lattice vectors
either because of symmetry considerations or because of photon polarizations.
By noticing the reciprocal lattice vectors at which the signal from a particular band is absent or very weak,
it may be possible to identify the $t_{2g}$ components which contribute dominantly to that band.  This orbital information
strongly constrains the band model.

In the sudden approximation
\bea
\langle \bm{p} \vert \Psi_{n\bm{k_\parallel}} \rangle &=& \delta_{\bm{k_\parallel+G_\parallel,p_\parallel}} \sum_j a^{(n)}_j(\bm{k_\parallel})  \nonumber  \\
&\times& \int \bm{dr}dz e^{-i[\lp \bm{G_\parallel r}+p_z z \rp}  \xi_j(\bm{r},z).
\label{M_sudden_approx}
\eea
Here $\bm{G_\parallel}$ is the surface-plane projection of
a reciprocal lattice vector, and
$\xi_j$ are the $t_{2g}$ basis functions given by Eq.(\ref{basis_functions}) for the conduction band
initial wavefunction: $\Psi_{n\bm{k_\parallel}}=\exp\lp i\bm{k_\parallel r} \rp\sum_j a^{(n)}_j \xi_j$.
The $\delta$-function in Eq.(\ref{M_sudden_approx}) reflects the conservation of the in-plane crystal momentum in the photon assisted scattering process of the electron.
\begin{figure}
	\centering
	\includegraphics[width=\columnwidth]{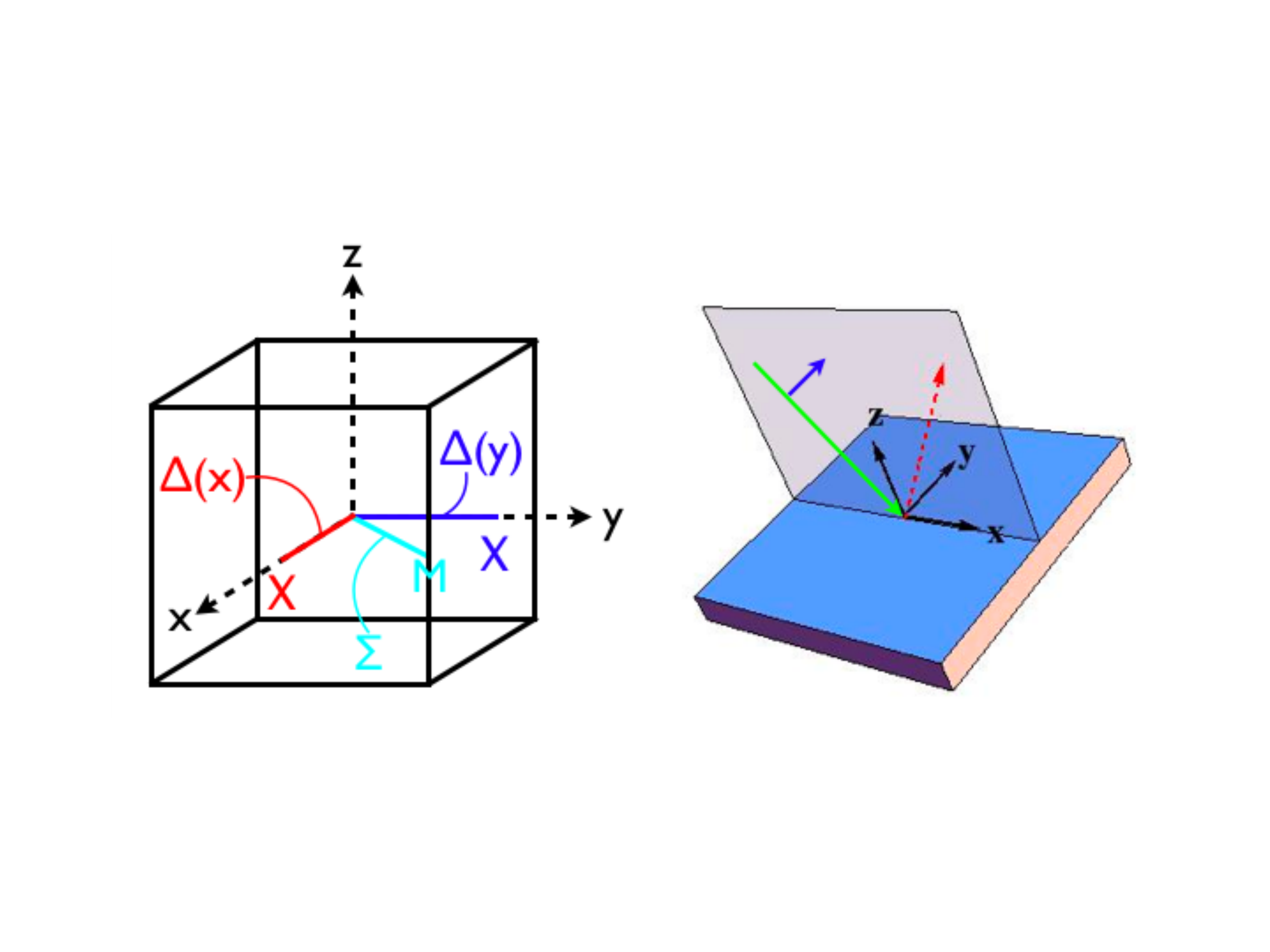}
	\caption[]{(Color online) Left: High symmetry points and lines in the Brillouin Zone (BZ) of a simple cubic lattice.
    Right: Experimental geometry used in simulations of ARPES data in the [10] BZ.  The photon source (green), with polarization in the xz-plane (blue),
    excites an electron to a high energy state that is emitted towards the detector (red dashed).  Although the experimental geometry is unchanged
    by a reflection through the xz-plane, the yz and xy orbitals are odd under this operation.  This leads to measurement of only the zx band in this measurement.}
	\label{fig:mirrorPlanes}
\end{figure}

\begin{figure}
	\centering
	\includegraphics[width=0.7\columnwidth]{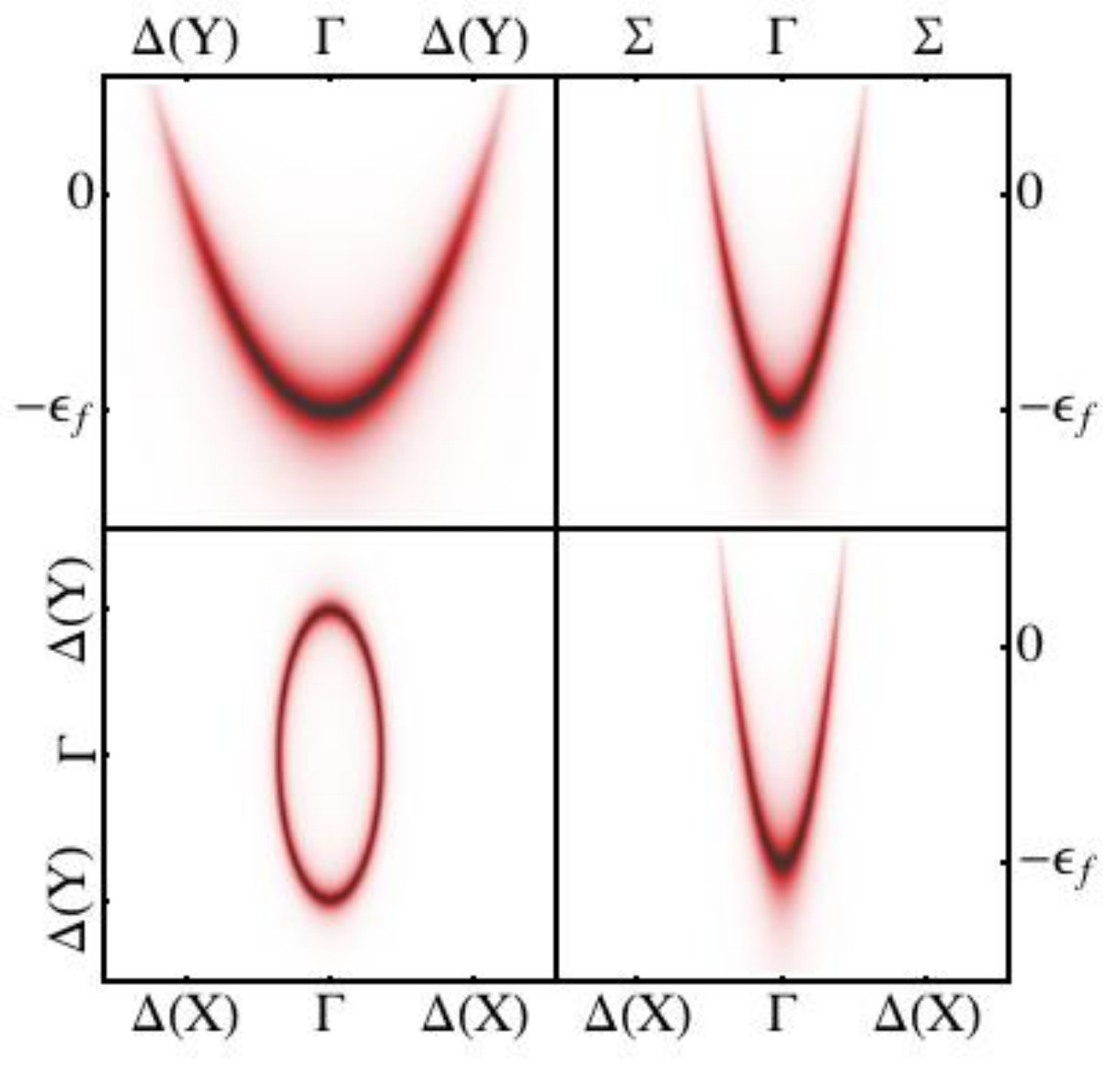}
	\caption[]{(Color online) Simulated ARPES signal in the [10] BZ for a temperature of $120K$, $L/M=1/8$, $\Delta_{\ty{SO}}=0$, and  $N=0$.
    The resolution for all ARPES simulations was set to $10$meV.  Only a single elliptical FS cross-section is seen (bottom left).
    Energy distribution curves (EDCs) have been included along several high symmetry directions - in this case showing only a single band associated with the $xz$ basis state.}
	\label{fig:arpes_0}
\end{figure}

We illustrate the usefulness of the matrix element effect by considering $M$ for $\bm{G}=\bm{0}$ ([00] BZ) and for $\bm{G}$  along the $x$-axis ([10] BZ).
In the first case $M$ always vanishes since all $\xi_j$'s are odd with respect to reflection off either the z-x or the z-y plane.
There is therefore no ARPES signal in the [00] BZ for $t_{2g}$ conduction band states.
For the [10] BZ
\be
M^{[01]} \propto \int \bm{dr}dz e^{-i\lp G_x x+ p_z z \rp}  Y(\bm{r},z).
\ee
Contributions from other $t_{2g}$ components of $\Psi_i$ vanish because of their reflection symmetry in the x-z mirror plane (see Fig.\ref{fig:mirrorPlanes}). Therefore only
wave functions containing a $Y$ orbital will be detected in this case. Recent experiments\cite{Chang,Meevasana} on bulk SrTiO{$_3$} find a single (doubly degenerate)
band for $k_z=0$ in the [01] BZ in the cubic as well as in the tetragonal phase (see section \ref{sec_STO}).
The Hamiltonian described by Eqs.(\ref{DeltaT}---\ref{h}) then indicates that the $\cN$'s
and $\Delta_{\ty{SO}}$ must be sufficiently small so that any hybridization between the $t_{2g}$-orbitals is negligible. (see Fig.\ref{fig:arpes_0}).

\begin{figure}
	\centering
	\includegraphics[width=\columnwidth]{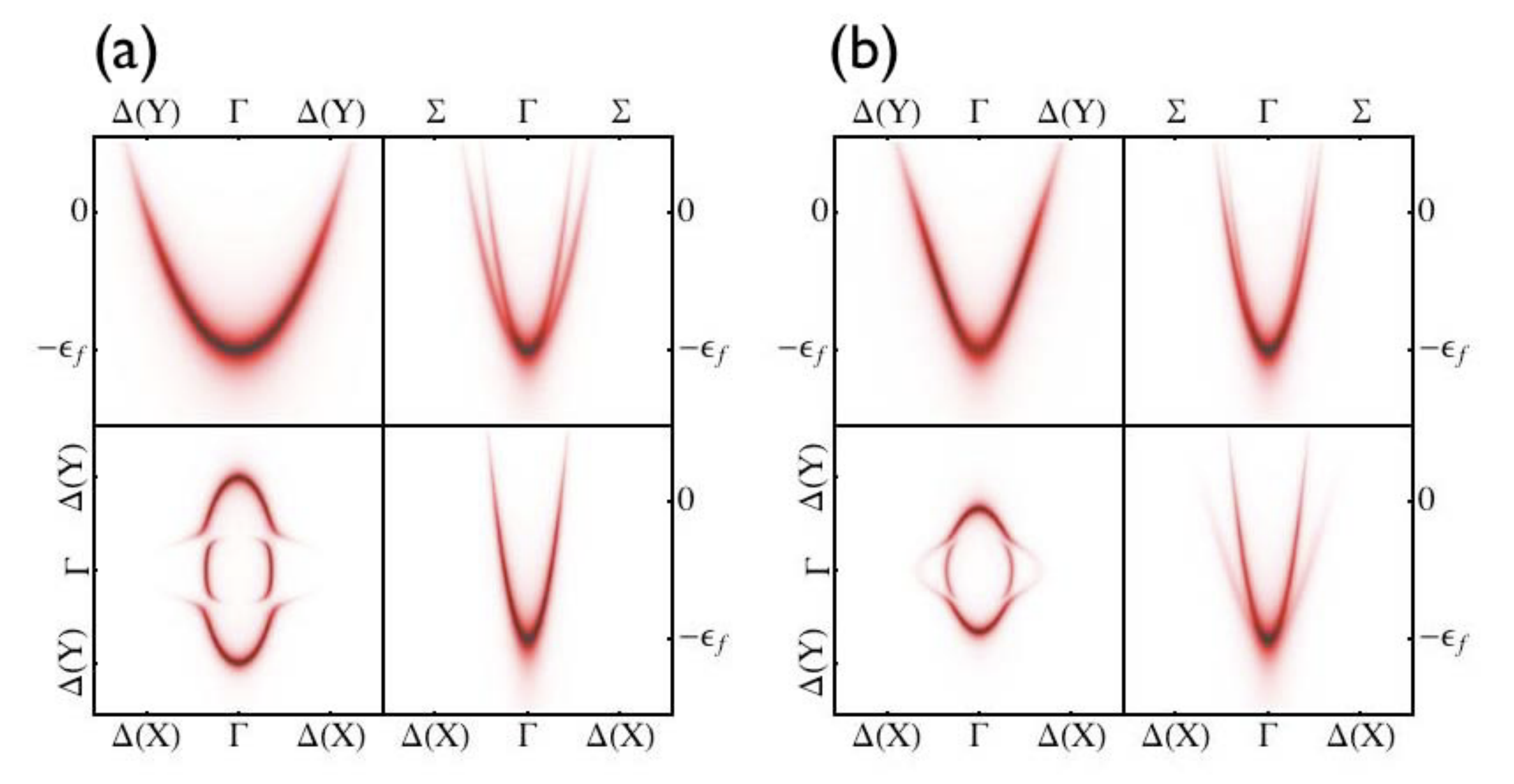}
	\caption[]{(Color online)	Comparison of d-orbital mixing. Simulated ARPES signal for a temperature of $120K$, $L/M=1/8$, and  $\Delta_{\ty T}=0$.
    (a) For $N/M=3$ and $\Delta_{\ty{SO}}=0$ the hybridization is most pronounced along $\Sigma$ but unseen in the EDCs along $\Delta(X)$, and $\Delta(Y)$.
    (b) For $N/M=0$ and $\Delta_{\ty{SO}}=3 \epsilon_F$ one band has moved above the Fermi energy.  The hybridization between the
    basis states is seen in the EDCs along all directions.  This experimental feature can be attributed to the lack of a preferential direction of the SO interaction. }
	\label{fig:arpes_N}
\end{figure}

As evident from the Hamiltonian (\ref{h_ij}) and illustrated in Fig.\ref{fig:arpes_N} the $t_{2g}$ d-orbitals are hybridized by $N$.
The influence of $N$ is most pronounced along the main diagonals. For example in the $[110]$ direction $N$ induces a momentum dependent gap of $2Nk^2$.

Spin-orbit interactions will also mix the $t_{2g}$ d-orbitals however in contrast to the $N \ne 0$ scenario they have no preferential direction.
When the SO splitting is larger than the Fermi energy, the ARPES spectrum along the $\Sigma$ direction
is similar to the spectrum in the $N \ne 0$, $\Delta_{\ty{SO}}=0$ case.
However, unlike the $N\ne0$ case
the photoemission spectrum is also altered along the $\hat{\bm{x}}$ and $\hat{\bm{y}}$ directions.
This is evident along $k_x$ in the simulated ARPES data of Fig.\ref{fig:arpes_N}, where the induced hybridization of the basis functions
cause the previously dark band to become visible.
\begin{figure}
	\centering
	\includegraphics[width=\columnwidth]{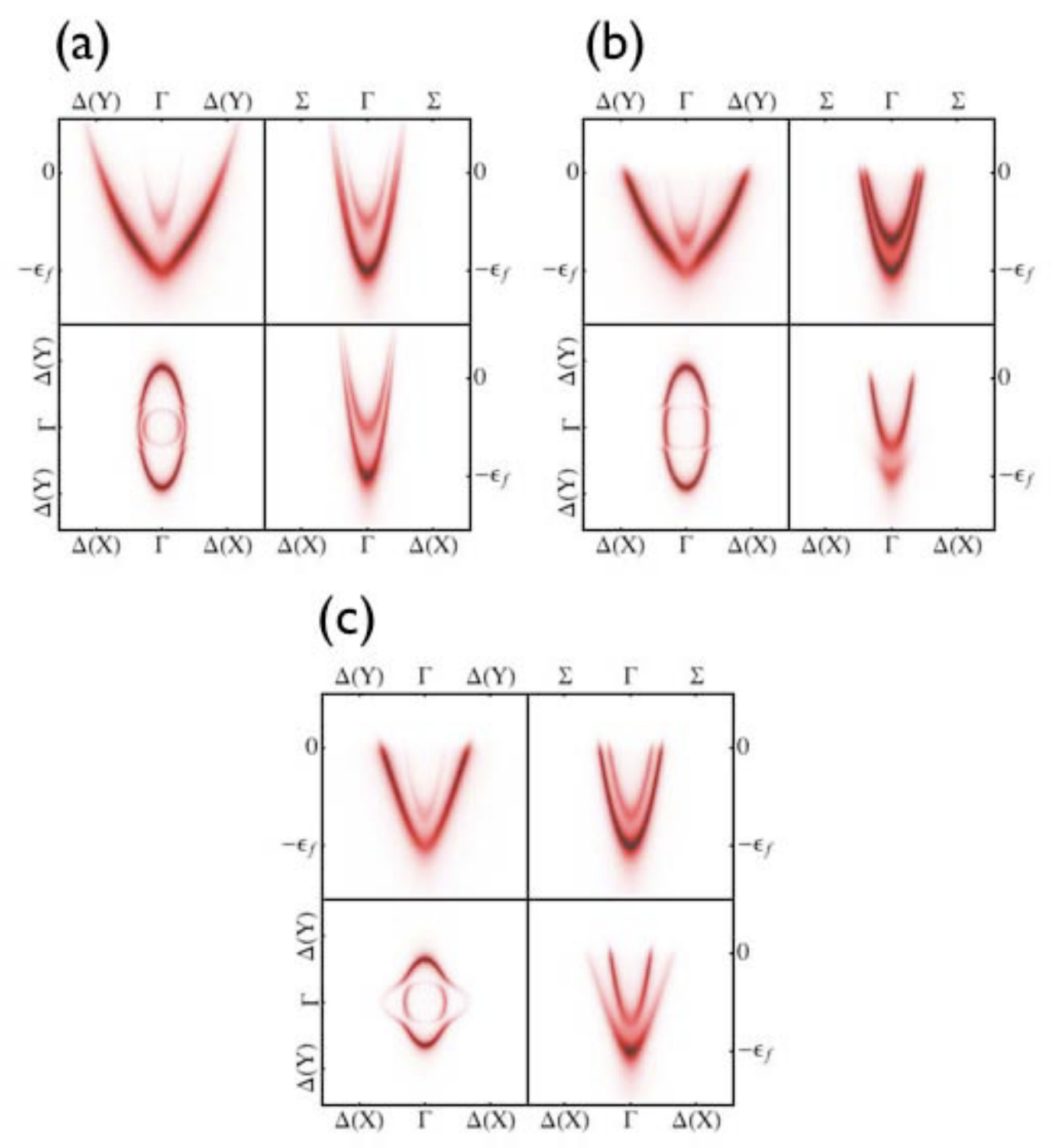}
	\caption[]{(Color online) Simulated ARPES measurement for the [10] BZ for different values of $\Delta_{\ty{SO}}$ and $\Delta_{\ty T}$.
    $L/M=1/8$, and $N=0$ in all figures.
    (a) $T=120K$, $\Delta_{\ty{SO}}= 0.5 \epsilon_{\ty F}$, $\Delta_{\ty T}=0$.  As evident from all EDCs the SO splitting hybridizes the basis states.
    If the temperature is lowered, inducing a structural phase transition that is large, one band moves above the Fermi energy.
    This relatively weak hybridization is seen in the Fermi surface (FS) of (b) where $T=20K$, $\Delta_{\ty{SO}}= 0.5 \epsilon_{\ty F}$, $\Delta_{\ty T}= 3 \epsilon_{\ty F}$.
    In contrast, for $T=20K$, $\Delta_{\ty{SO}}= 3 \epsilon_{\ty F}$, $\Delta_{\ty T}=0.5 \epsilon_{\ty F}$ the strong hybridization of the basis states leads to a more symmetric FS (c).
    This feature is also seen by comparing the EDCs of (b) and (c).}
	\label{fig:arpes_SO}
\end{figure}

\begin{figure}
	\centering
	\includegraphics[width=\columnwidth]{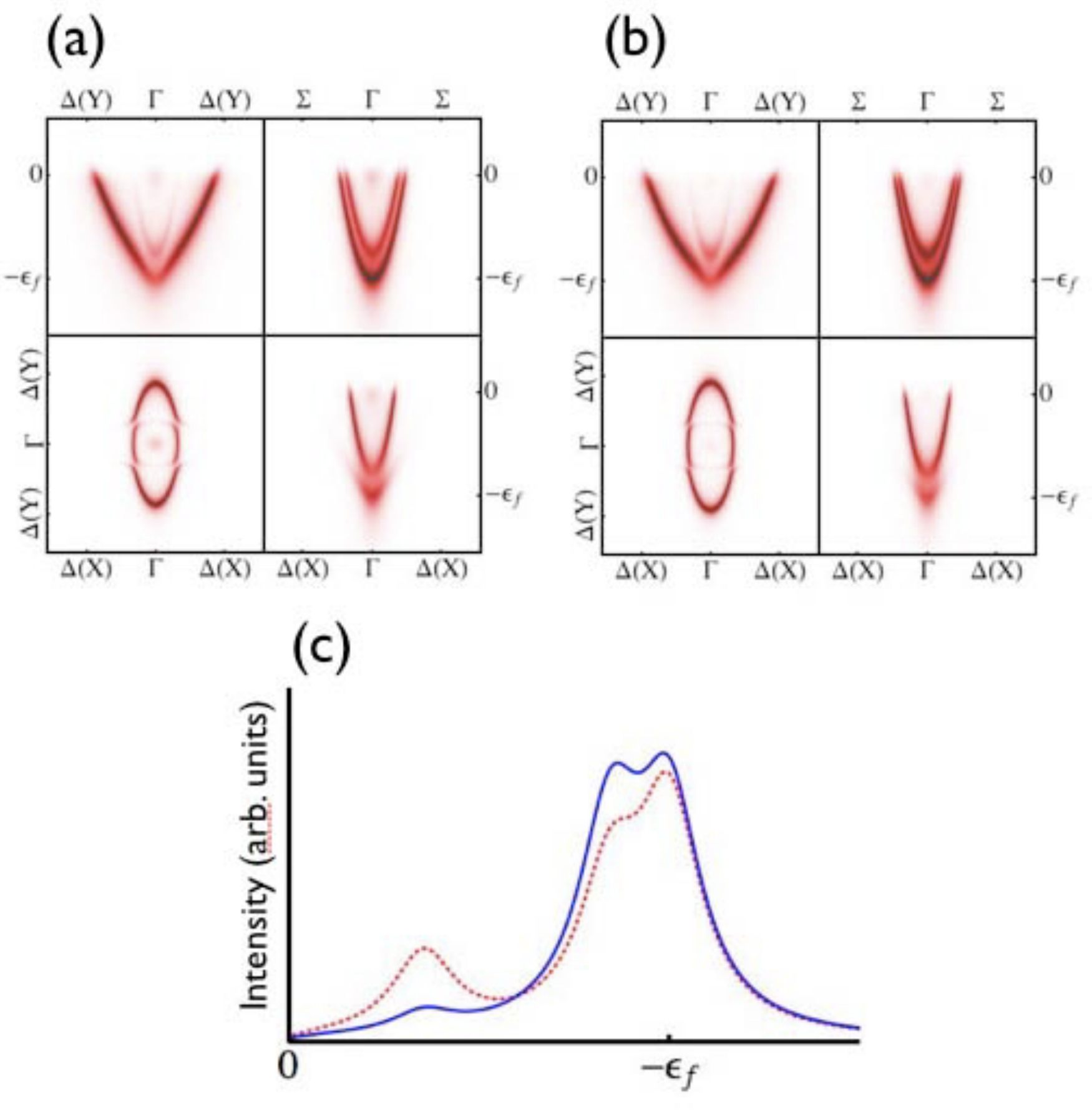}
	\caption[]{(Color online)  When $\Delta_{\ty{SO}}$ and $\Delta_{\ty T}$ are both small, although all three bands can be seen directly in a measurement of the $[10]$ BZ it can
    be unclear how to extract these parameters.  This is seen by comparing (a) where $\Delta_{\ty{SO}}= .5 \epsilon_{\ty F}$. $\Delta_{\ty T}=0.3\epsilon_{\ty F}$ with (b) where
    $\Delta_{\ty{SO}}= 0.3 \epsilon_{\ty F}$. $\Delta_{\ty T}=0.5\epsilon_{\ty F}$.  A careful analysis of the EDC at the $\Gamma$ point can distinguish between the two scenarios.
    This is shown in (c) where the $\Gamma$ point EDC is shown for (a) in dashed-red and for (b) in blue. }
	\label{gammast}
\end{figure}

The SO and tetragonal energies $\Delta_{\ty{SO}}$ and $\Delta_{\ty T}$ determine the band splitting at the $\Gamma$ point (see Eq.(\ref{E_tetragonal})).
In the cubic phase the value of $\Delta_{\ty{SO}}$ can be extracted directly from an ARPES measurement in the $[10]$ surface BZ (see Fig. \ref{fig:arpes_SO}a).
This simple picture is complicated in the tetragonal phase.
The case where $\Delta_{\ty{SO}}>\epsilon_{\ty F}$ and $\Delta_{\ty T}<\epsilon_{\ty F}$ is readily
distinguished from the opposite limit by analysis of the dispersion along $k_x$.  As evident from Figs. \ref{fig:arpes_SO}b and \ref{fig:arpes_SO}c
only in the case where $\Delta_{\ty{SO}}>\epsilon_F$, is the dark weakly dispersive band visible away from the $\Gamma$ point.
When both $\Delta_{\ty{SO}}$ and $\Delta_{\ty T}$ are less than the Fermi energy,
the Energy Distribution Curves (EDCs) at the $\Gamma$ point can be used to distinguish between the two cases.
This can be seen in Fig. \ref{gammast}.

\subsection{Magnetic Oscillations       \label{sec_magnetic_oscillations}}

Magnetic oscillations in various physical properties such as the conductivity (Shubnikov - de Haas effect) and the magnetic susceptibility (de Haas - van Alfen effect)
provide invaluable information on the band structure of solids\cite{Ashcroft, Marder, Kubler,Singleton}.
The frequency of the oscillations $F$ is related to the extremal cross-sectional area $A_{k}$
of the Fermi surface in a plane perpendicular to the magnetic field through the Onsager relation $ F = \phi_{0}A_k/4 \pi^2$.
Here $\phi_0=hc/e$ is the magnetic flux quantum. Measuring $F$ as a function of charge density, magnetic field orientation, and temperature
also makes it possible in principle to determine all the phenomenological Hamiltonian parameters.

In the naive picture of three ellipsoidal decoupled d-bands the cross sectional areas are simply given by ellipses.
However this oversimplified scenario breaks down for any realistic system due to the hybridization of the d-orbitals by $N$, and by the SO interactions.
Avoided crossings of the overlapping energy bands then result in more complicated energy surfaces.

To illustrate the variety of possible shapes of electron pockets we consider a simple case with a small but finite band mixing (e.g. $N \gtrsim 0$).
$\Delta_{\ty T}/\epsilon_{\ty F}=0.5$.
\begin{figure}
	\centering
	\includegraphics[width=1\linewidth ]{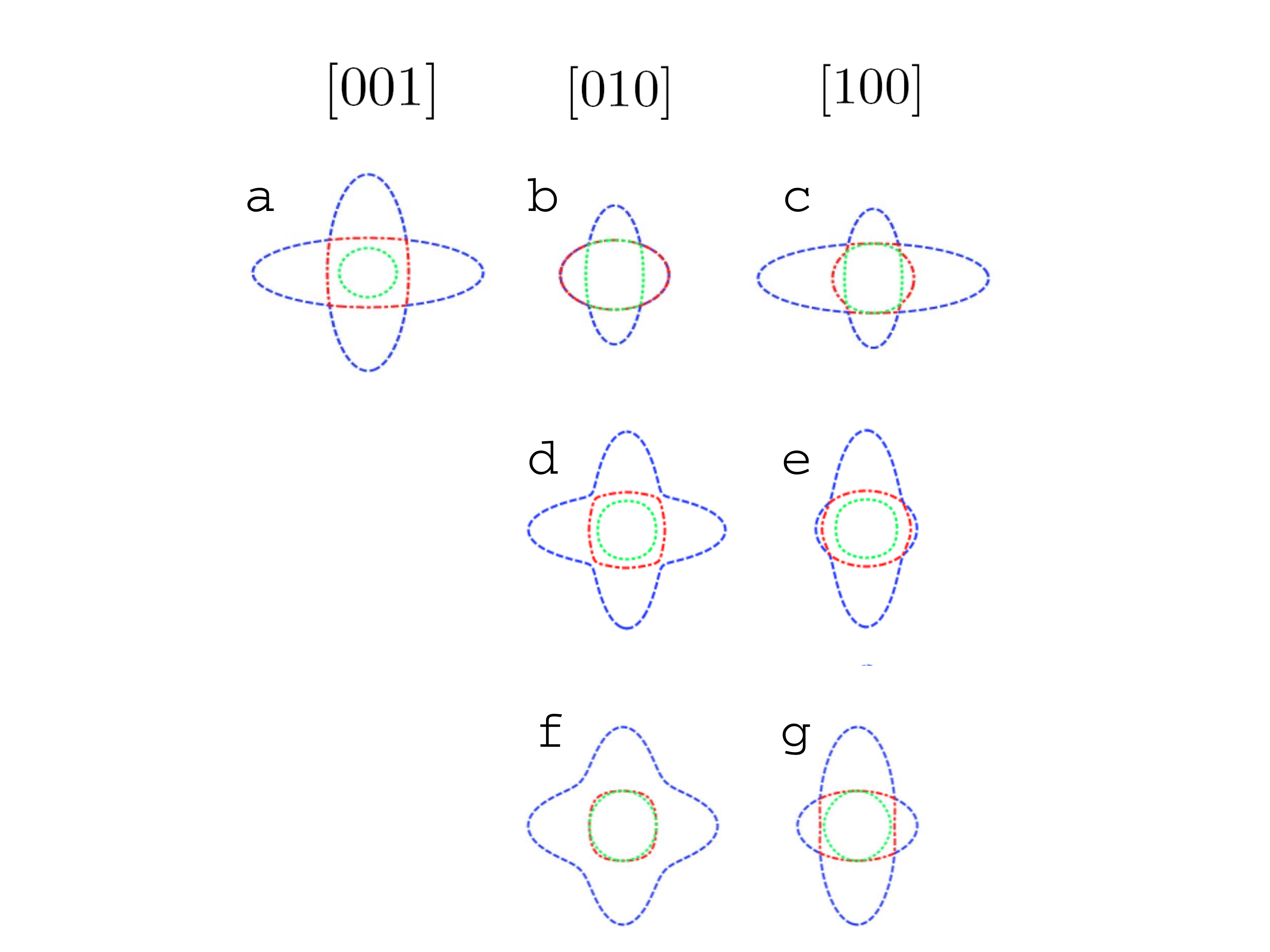}
	\caption[]{(Color online) Extremal cross sectional areas for magnetic field oriented along [001] (left) [010] (center) and [100] (right).
    The extremal orbits have been organized by size from largest to smallest and shown as dashed-blue, dot-dashed-red, and dotted-green, respectively.
    Top row (a-c) corresponds to $\Delta_{\ty T}=0.5\epsilon_{\ty F}$, middle row (d,e) corresponds to $\Delta_{\ty{SO}}=0.5\epsilon_{\ty F}$, and
    bottom row corresponds to $N=0.5M$. }
	\label{fig:extremalAk}
\end{figure}

The cross sectional areas for three
high symmetry directions of the magnetic field are depicted in the top row of Fig.\ref{fig:extremalAk} for the tetragonal phase.
As $\Delta_{\ty T}$ increases the most energetic band is gradually depleted
and the electronic charge is redistributed amongst the other two Fermi pockets. Eventually for
$ \Delta_{\ty T}/\epsilon_{\ty F} >1-\textrm{min}(L/M,M/L)$
there is no band crossing between the $xy$-band and the other two bands.

Avoided crossings in the cubic phase result in non-elliptical cross-sections as well.  The extremal cross-sectional areas along high symmetry directions are depicted in
Fig.\ref{fig:extremalAk} for $\Delta_{\ty{SO}}/\epsilon_{\ty F}=0.5$ (center row) and for $N/M=0.5$ (bottom row).

Our discussion ignores the possibility of multiple domains
in the distorted state, and neglects magnetic breakdown.
The latter is likely present in
magnetic oscillation measurements on these materials because of the close approaches between extremal cross-sections \cite{Chambers}
belonging to different bands.

\section{$\textrm{SrTiO}_3$      \label{sec_STO}}

Bulk STO is a band insulator with an energy gap of $3.2$ eV. By chemical substitution of the $Ti$ or the $Sr$ atoms or by introducing
oxygen vacancies it is possible to electron dope the system with a high level of precision.
STO has cubic symmetry at room temperature, however at 105K it undergoes a antiferrodistortive structural transition to a tetragonal phase.
Below the critical temperature neighboring $TiO_6$ octahedras continuously rotate in opposite directions by an angle
of up to a few degrees.

Although STO has been studied for many years, there are only a few experimental results that can shed light on the structure of its conduction bands.
We therefore resort to a 5-parameter model in which the Hamiltonian
is parameterized by $\Delta_{\ty{SO}},\Delta_{\ty{T}},M,N$, and $L$, {\em i.e.} $h$ is approximated by its cubic phase form.

The experiments
that do exist appear to partially contradict one another.
Based on Raman spectroscopy and Shubnikov de-Hass measurements Uwe {\em et. al.}\cite{UweRaman,UweSdH} concluded that $\Delta_{\ty{SO}}\approx 18$meV and $\Delta_{\ty T}\approx 1.5$meV.
On the contrary, Chang {\em et. al.}\cite{Chang} using ARPES do not observe a SO induced gap at the zone center and conclude that $\Delta_{\ty T} \approx -25$meV.

Supporting evidence for the smallness of $\Delta_{\ty{SO}}$  is provided by the matrix element effect.
Experiments\cite{Chang} observe only, what should be according to our matrix element analysis, the X orbital in the [10] BZ and the Y orbital in the [01] BZ.
As explained in section \ref{sec_ARPES}, the lack of hybridization between $t_{2g}$ orbitals implies that both $\Delta_{\ty{SO}}$ and $N$ are very small.
Additional proof that $N\ll M,L$ is provided the ARPES EDCs which reveal no special features of the energy along $\Sigma$.
In addition, these curves yield values for the effective masses from which it follows that
\be
M \approx 0.84   \ , \ L \approx 0.14.        \label{ML}
\ee

\begin{figure}
	\centering
	\includegraphics[width=\linewidth]{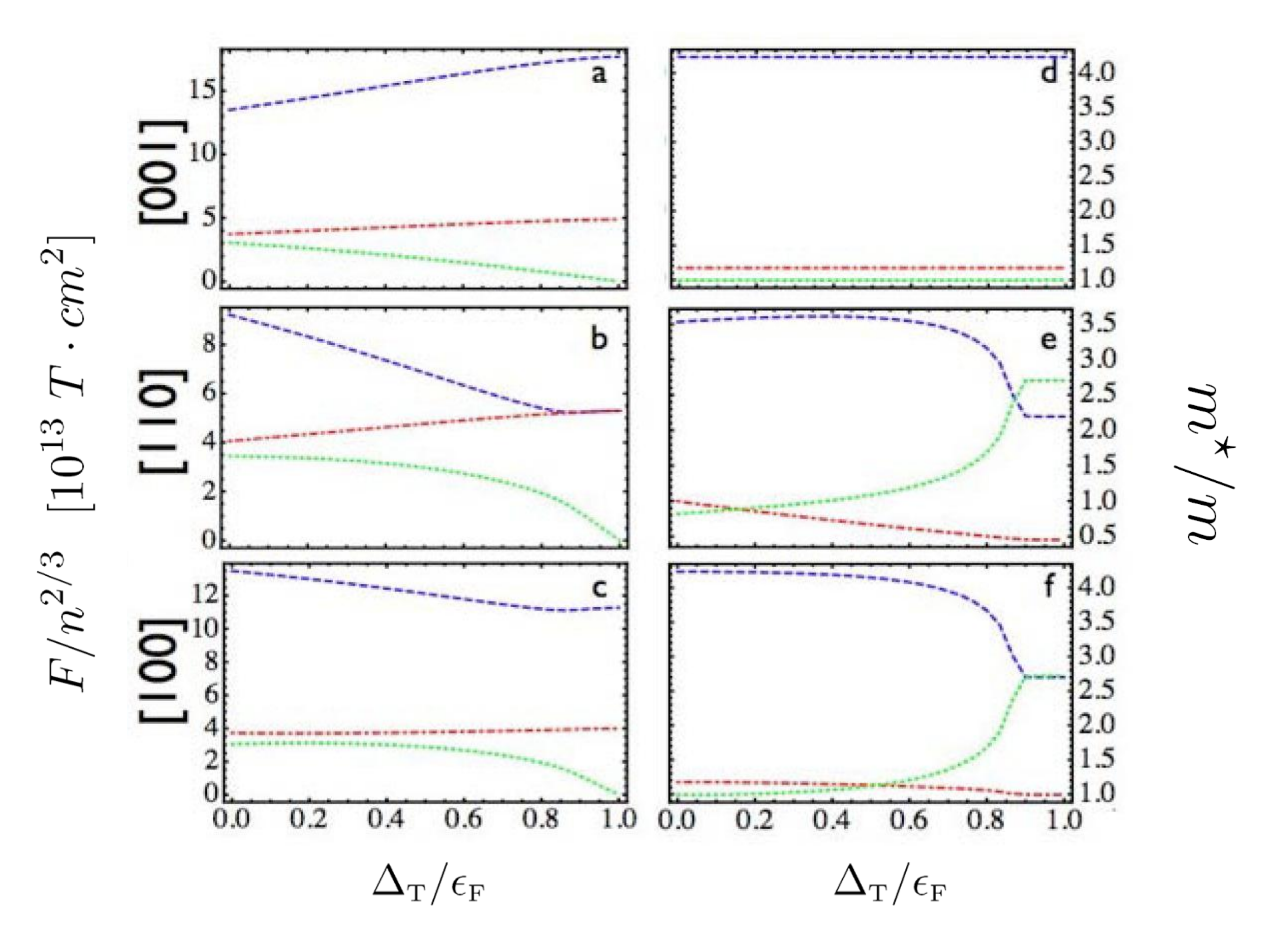}
	\caption[]{(Color online) Dependence of magnetic oscillation frequency and cyclotron mass on $\Delta_{\ty T}$. Here we set $L/M=1/8$, $N=0$, and $\Delta_{\ty{SO}}=0$.
	(a,b,c): Scaled SdH frequency for magnetic fields along $[001]$, $[110]$, and $[100]$ as a function of $\Delta_{t}$.
	(d,e,f): Cyclotron mass as a function of $\Delta_{\ty T}$.}
	\label{fig:SdH_T}
\end{figure}
\begin{figure}
	\centering
	\includegraphics[width=\linewidth]{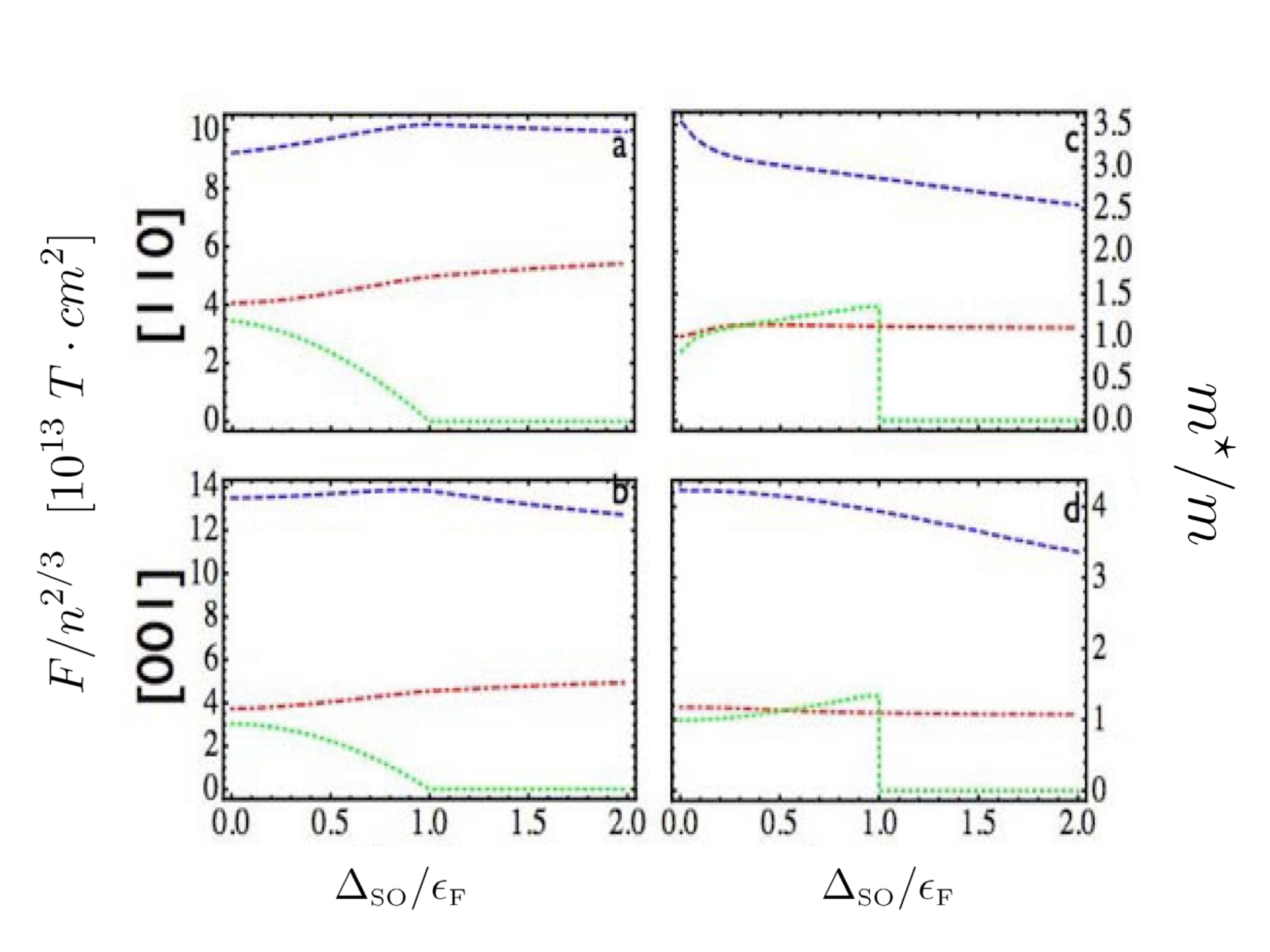}
	\caption[]{(Color online) Dependence of magnetic oscillation frequency and cyclotron mass on $\Delta_{\ty{SO}}$. Here $L/M=1/8$, $N=0$ and $\Delta_{\ty T}=0$.
	(a,b): Scaled SdH frequency for magnetic fields along $[001]$, $[110]$, and $[100]$ as a function of $\Delta_{\ty{SO}}$.
	(c,d): Cyclotron mass as a function of $\Delta_{\ty{SO}}$.
	}
	\label{fig:SdH_s}
\end{figure}

Raman spectroscopy measurements\cite{UweRaman} find energy gaps of approximately 2meV and 18 meV between conduction bands at the $\Gamma$ point suggesting that $\Delta_{\ty T}$ and $\Delta_{\ty{SO}}$
have very different magnitudes.  The larger of the two scales can be identified as tetragonal or spin-orbit from the dependence of magnetic oscillation frequency $F$ and cyclotron mass $m^\star$
on density and field orientation.
Fig.\ref{fig:SdH_T} depicts the dependence of $F$ and $m^\star$ on density and on $\Delta_{\ty T}$.
The dependence can be expressed through a single parameter $\Delta_{\ty T}/\epsilon_{\ty F}$ if $F$ is scaled with $n^{2/3}$ where
$n$ is the electronic density. Similar graphs are given in Fig.\ref{fig:SdH_s} for a scenario in which $\Delta_{\ty T} \ll \Delta_{\ty{SO}}$. The different trends
of $F$ and $m^\star$ as a function of density clearly distinguishes between the $\Delta_{\ty T} \gg \Delta_{\ty{SO}}$ scenario and its opposite counterpart.

\section{Summary    \label{sec_summary}}

$d^0$ Perovskites have played a centeral role in various areas of solid state physics and are now emerging as important building blocks
for oxide-based hetro-structures. In this work we used the $\bm{k \cdot p}$ theory to construct the general low energy theory for the conduction bands of these materials
both in the cubic and in the tetragonal phases. We then employed the theory to estimate the Hamiltonian parameters for STO.

Our work emphasizes the need for additional experimental data on the electronic band structure of Perovskites.
Even for STO, by far the most studied $d^0$ Perovskite,
existing experimental data is insufficient to uniquely determine the values of
band parameters that will, for example, control the character of the two-dimensional
electron systems formed by $\delta$-doping.

In the past few years much effort has been devoted to fabricating oxide-based hetro-structures.
Our model for the electronic structure of the bulk is a first step towards modeling these complex systems.

This work was supported by Welch Foundation Grant F1473.
We acknowledge helpful conversations with Jim Allen, Young Jun Chang, Harold Hwang, Worawat Meevasana, and Susanne Stemmer.

\appendix

\section{$\bm{k \cdot p}$ Hamiltonian for the tetragonal phase  \label{app_H}}

The momentum dependent part of the effective hamiltonian $h$ for a $d^0$ perovskite is given by Eq.(\ref{h_ij}).
In this appendix we use group theory methods to express $h$ in terms of a small number of phenomenological parameters\cite{Dresselhaus}.

The calculation of $h$ involves the evaluation of matrix elements of the form $\me{\phi}{\bm{k \cdot p}}{\psi_j}$.
Here $\psi_j$ is a basis function of the $t_{2g}$ manifold, $\bm{p}$ is the momentum operator, and $\phi$ is a state
outside of the $t_{2g}$ manifold.
At the cubic to tetragonal phase transition the symmetry at the zone center reduces from
$O_h$ to $D_{4h}$.
Correspondingly, at the phase transition the three $\psi_j$'s change their transformation properties $\Gamma_{25}^+ \to \Gamma_{4}^+ + \Gamma_{5}^+$.
The three components of the momentum operator $\bm{p}$, that transform as a single irrep ($\Gamma_{15}^-$) in the cubic phase split:
$p_x,p_y \in \Gamma_5^-$ whereas $p_z \in \Gamma_2^-$.

The values of the matrix elements vary smoothly across the structural transition.
To emphasize the relation between the two symmetries we label the matrix
elements in the tetragonal phase with a subscript that corresponds to the irrep of $\phi$ in the cubic phase and a superscript that corresponds to its
irrep in the tetragonal phase . For example,
$B^5_{15}$ is associated with a basis function that evolved from $\Gamma_{15}$ in the cubic phase to $\Gamma_5$ in the tetragonal one.

We first consider the $\Gamma_4^+$ band.
The intermediate states are
\bea
\Gamma_5^- \otimes \Gamma_4^+ &=&   \Gamma_5^-            \nonumber \\
\Gamma_2^- \otimes \Gamma_4^+ &=&   \Gamma_3^-.
\eea
Denoting
\bea
\me{\Gamma_{5x}^-}{(p_x,p_y)}{\Gamma_4^+} &=& \left\{
     \begin{array}{c}
       B_{15}^5 (0,1) \\
       B_{25}^5 (0,1)
     \end{array}  \right.       \nonumber \\
\me{\Gamma_{5y}^-}{(p_x,p_y)}{\Gamma_4^+} &=& \left\{
     \begin{array}{c}
       B_{15}^5 (1,0) \\
       -B_{25}^5 (1,0)
     \end{array} \right.        \nonumber \\
\me{\Gamma_{3}^-}{p_z}{\Gamma_4^+} &=& \left\{
\begin{array}{c}
  B_{2}^3 \\
  B_{12}^3
\end{array}  \right. ,
\eea
we find that
\be
h_{zz}=\cM_B(k_x^2+k_y^2) + \cL_B k_z^2      \label{Ek_Gamma4p}
\ee
where the two real phenomenological parameters are given by
\be
\cM_B = \frac{1}{2m} + \frac{1}{m^2}\sum_{n' \in \Gamma_5^-} \frac{|B_{n'}^5|^2}{E^{\Gamma_4^+}(0)-E_{n'}^{\Gamma_5^-}(0)},
\ee
and
\be
\cL_B = \frac{1}{2m} + \frac{1}{m^2}\sum_{n' \in \Gamma_3^-} \frac{|B_{n'}^3|^2}{E^{\Gamma_4^+}(0)-E_{n'}^{\Gamma_3^-}(0)}.
\ee

We now turn to the $\Gamma_5^+$ band. The intermediate states are
\bea
\Gamma_5^- \otimes \Gamma_5^+ &=&   \Gamma_1^- + \Gamma_2^- + \Gamma_3^- + \Gamma_4^-           \nonumber \\
\Gamma_2^- \otimes \Gamma_5^+ &=&   \Gamma_5^-.
\eea
Following similar steps to those taken above we obtain expression (\ref{h}). The k-dependent Hamiltonian $h$ depends on six real parameters and a single
complex parameter $\cN_{BC}$. The $\bm{k \cdot p}$ Hamiltonian in the cubic phase can easily be obtained from its tetragonal counterpart by disregarding
the subscripts of the phenomenological parameters; for example by associating with $\cL_B$ and $\cL_C$ a single parameter $\cL$.

\end{document}